# SPD deformation of pearlitic, bainitic and martensitic steels

M. W. Kapp[1,*], A. Hohenwarter[2], A. Bachmaier[1], T. Müller[3], R. Pippan[1]

[1] Erich-Schmid-Institute of Materials Science, Jahnstraße 12, 8700 Leoben, Austria
[2] Department of Materials Science, Montanuniversität Leoben, Jahnstraße 12, 8700 Leoben, Austria
[3] Anton Paar GmbH, Anton-Paar-Straße 20, 8054 Graz, Austria
*corresponding author, marlene.kapp@oeaw.ac.at

Abstract

The deformation behavior of nearly fully pearlitic, bainitic and martensitic steels during severe plastic deformation is summarized in this paper. Despite their significantly different yield stresses and their microstructures, their hardening behavior during SPD is similar. Due to the enormous hardening capacity the SPD deformation is limited by the strength of the tool materials. The microstructure at the obtainable limit of strain are quite similar, which is a nanocrystalline structure in the order of 10 nm, dependent on the obtainable strain. The nanograins are partially supersaturated with carbon and the grain boundaries are stabilized by carbon. Another characteristic feature is the anisotropy in grain shape which results in an anisotropy of strength, ductility and fracture toughness. The results are important for the development of ultra-strong materials and essential for this type of steels which are frequently used for application where the behavior under rolling contact and sliding contact is important.

## 1 Motivation

Carbon-based steels, such as pearlitic, bainitic and martensitic ones, have played an important role in the industrial development of our society and are widely used in an enormous number of applications. Among application fields, such in construction, civil or mechanical engineering, these "simple" steels, mainly consisting of iron and carbon, are a backbone of the railway transportation sector. In this field particularly pearlitic and bainitic qualities are used for rails and crossings. The degradation and lifetime of rails is controlled by the typical stress state. In certain service situations high contact forces combined with shear forces act in the contact zone between rail and wheel and lead to so-called rolling contact fatigue (RCF) [1,2]. Typical RCF related defects, such as headchecks and squats can arise and endanger the traffic safety [3].

Field investigations showed that these conditions also lead to drastic material modifications on the rail surface with widely unknown mechanical property changes, which will dictate the RCF-behavior [4,5]. It was found that by means of Severe Plastic Deformation (SPD) such structures can be experimentally simulated to study the microstructural evolution and connected mechanical changes under better experimental control providing well defined strains and environmental conditions [6,7]. Therefore, beside other emerging application fields of SPD-processed materials, for example as medical implants, the modification of rail steels by high pressure torsion (HPT) represents an indirect industrial application field. This is not only restricted to rail steels but can find a more general application where high contact pressure and shear forces dominate the stress state and material degradation. For instance, in bearings or gear wheels often consisting of martensitic qualities, similar damage events and material modification such as the occurrence of white etching layers can be found, which threaten the safe operation of machines [8,9]. The near surface deformation during rolling contact of these components can be considered as a SPD surface treatment.

The importance of the microstructural changes of such steels for the understanding of the material strengthening, but also the connected changes in ductility and fracture resistance motivated us to give an overview on the most important structural features and changes of pearlitic, bainitic and martensitic steels during SPD.

## 2 Structural evolution of steels during SPD

### 2.1 Pearlitic steels during wire drawing

Although wire drawing has been practiced for hundreds of years [10], the theoretical framework to understand the underlying mechanical and metallurgical processes was just built in the past 60 years. To date, cold-drawn pearlitic steels belong to the best studied steels in SPD. Since the seminal work of Embury and Fisher in 1966 [11] on pearlitic steels, the evolution of structural features during wire drawing was extensively studied. The authors revealed decreasing ferrite cell size with smaller wire diameter, which is the source for the strengthening with increasing drawing strain. The flow stress, $\sigma_f$, can be predicted from the imposed drawing strain, $\varepsilon_d$, using a model based on Hall-Petch hardening with the initial ferrite spacing, $d_0$, and the empirically developed Hall-Petch constant, $k$:

$$\sigma_f = \sigma_i + \frac{k}{\sqrt{2}\sqrt{d_0}} \exp\left(\frac{\varepsilon_d}{4}\right) \quad \text{(Eq. 1)}$$

Because of the strong correlation between drawing strain and grain refinement [12], the strength can be directly linked to the square root of the ferrite lamellae spacing, $d$, with the mean shear modulus of ferrite and cementite, $G$, and the respective burgers vectors for a dislocation in cementite, $b_{cm}$, and ferrite, $b$ , and a constant, $\alpha$ = 0.7:

$$\sigma_f = \sigma_i + 0.35G\sqrt{\frac{b_{cm}}{d}} + \alpha G \frac{b}{d} \quad \text{(Eq. 2)}$$

Thus, the strength is controlled by the average slip distance of a dislocation in ferrite. This knowledge built the cornerstone of describing strengthening in severely drawn pearlite for a long time.

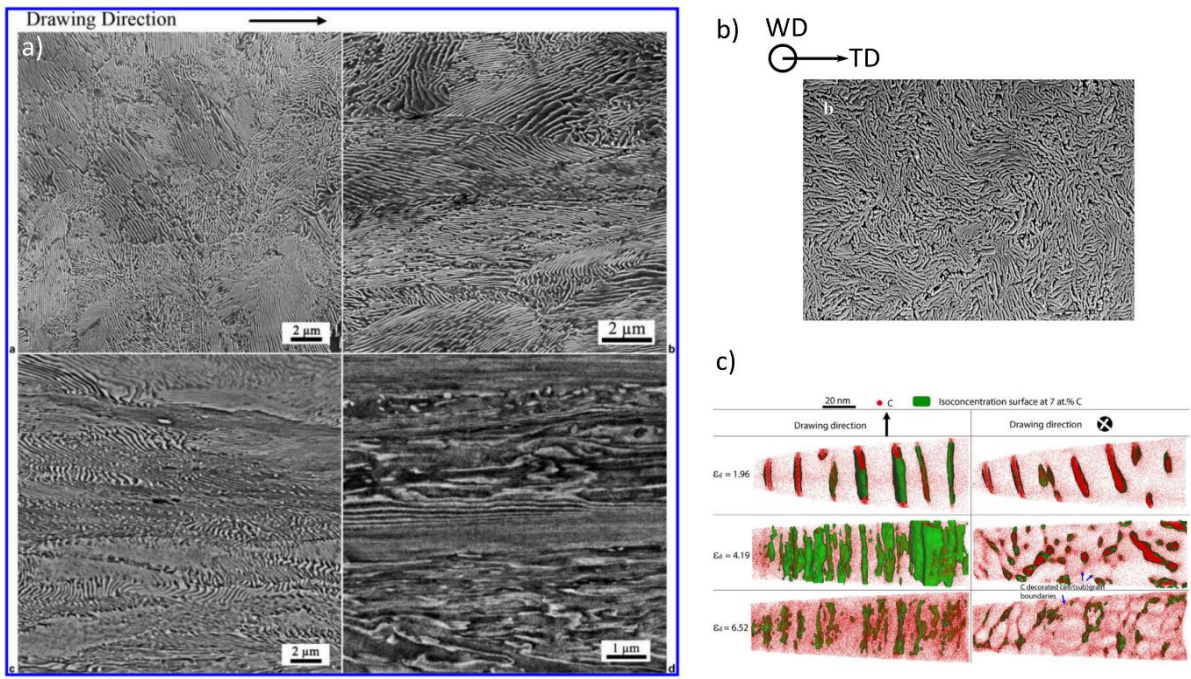

*Figure 1 a) Continuous realignment of cementite towards the drawing direction with increasing drawing strain proceeds faster in already preferentially aligned colonies ($\varepsilon_d$ = 0, 0.68, 1.51, 2.67) reprinted by permission of the publisher (Taylor & Francis Ltd, http://www.tandfonline.com) [13]. b) A curling structure develops perpendicular to the drawing direction (WD) in the*

*transversal direction (TD). Reprinted from [14] with permission from Elsevier. c) Evolution from an initially lamellar compound ($\varepsilon_d$=1.96) to a ferritic subgrain structure with carbon either supersaturated in the matrix or segregated to subgrain boundaries ($\varepsilon_d$=6.52), reprinted with permission from [15]. Copyright (2023) from the American Physical Society.*

The ferrite lamellae spacing prior to drawing is typically in the range of 100-300 nm and refines to values well below 10 nm for the highest drawing strains of $\varepsilon_d$ = 6.52 [15]. This goes along with an overall transition from a colony structure into a nanolamellar compound of ferrite and cementite aligned along the drawing direction [13,14,16–18]. Thereby, the cementite phase considerably affects the structural evolution. Though, its intermetallic nature would suggest a brittle behavior, it can plastically deform if it is thin enough [11,19,20]. Below 10 nm thickness it undergoes large plastic deformation [12,21,22] and thereby promotes structural refinement. During wire drawing the cementite may fragment or deform [23], which depends also on the initial alignment within the colony with respect to the drawing direction [14,22,24]. The refinement of the lamellae in colonies being aligned parallel to the drawing direction is most efficient (Figure 1a), as a realignment of the cementite is not required and the imposed strain mainly causes a thinning of the lamellae. In colonies where the lamellae are aligned perpendicular to the wire axis, the largest re-orientation needs to be realized and therefore the interlamellar spacing appears larger at moderate drawing strains. The zig-zag structure is a consequence of compressive straining, yielding and buckling of the cementite in case of fine pearlite. However, brittle fragmentation without considerable thinning of the cementite would be observed for coarser pearlite. The realignment of cementite lamellae is accompanied with the evolution of a curling structure (Figure 1b) perpendicular to the drawing axis [14,25]. Thus, efficient grain refinement and strengthening in such wires relies on an optimum starting structures, e.g. optimum rolling and cooling parameters during pre-manufacturing or lead patenting is necessary [26].

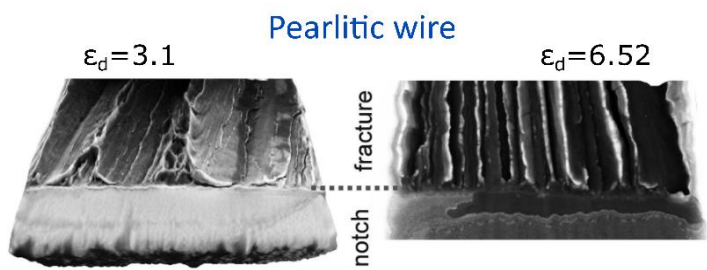

*Figure 2 Decohesion along the wire drawing axis for the lamellar architecture ($\varepsilon_d$=3.1) as well as for the nanograined one ($\varepsilon_d$=6.52) emphasizes the importance of the elongated interfaces for mechanical properties [27].*

In accordance to the knowledge gained over the past decades, the wire drawing procedure could be optimized so that diameter reductions down to 20 µm thickness are accessible, i.e. drawing strains of $\varepsilon_d$ = 6.52 [15]. This yields a tensile strength of 6.8 GPa and makes nanostructured pearlitic steels the strongest structural material to date. This exceptional strength is a consequence of the severe refinement down to 8 nm ferrite spacing (Figure 1c). Yet, at this stage of deformation the structural architecture has changed completely. Instead of a lamellar compound of ferrite and cementite, a ferritic subgrain structure with carbon either supersaturated in the matrix or segregated to boundaries develops. This change in structure is because severe deformation removes carbon from the cementite [15,17,28–32]. It thereby becomes off-stoichiometric already at lower drawing strains due to dislocation drag of carbon atoms [33], which then segregate to the interface[34]. Similarities in the fracture surfaces at $\varepsilon_d$ = 3 (nanolamellar composite) and $\varepsilon_d$ = 6.52 (nanograined ferrite) suggest, that still the interfaces in the drawing direction largely control mechanical properties (Figure 2) [27]. They are source for delamination upon failure, irrespective whether the cementite is stoichiometric [35], becoming non-stochiometric [36,37] or already transforms to a carbon film along the interface [27]. However, the transfer of a lamellar architecture to a single-phase nanocrystalline alloy reduces ductility and toughness [27,38], albeit deformability remains notably high with respect to its exceptional strength. It is therefore an outstanding example for SPD of steels and an inspiring role model for other methods of severe plastic deformation, such as HPT, equal channel angular pressing and accumulative roll bonding.

## 2.2 HPT deformation of pearlitic steels

Despite the differences in the deformation mode the structural changes, when pearlitic steels are deformed by quasi-constrained HPT, are very similar to wire drawing (Figure 3, first row) [39–42]: i) the initially random colony structure is transformed into a lamellar nanocomposite with the ferrite and cementite lamellae fully aligned along the shearing direction of the HPT disc ($\varepsilon_v$ = 11); ii) the realignment of the cementite proceeds faster in colonies already being aligned parallel to the HPT shearing direction (very fine lamellar spacing at $\varepsilon_v$ = 3), while in colonies being initially perpendicularly oriented, a zig-zag structure is present (coarse structure at $\varepsilon_v$ = 5); iii) the cementite becomes off-stoichiometric because of the dissolution of carbon atoms in the ferrite matrix [43]. Due to the high hydrostatic pressure compared to wire drawing, the initial thickness of cementite is less important for the successful processing.

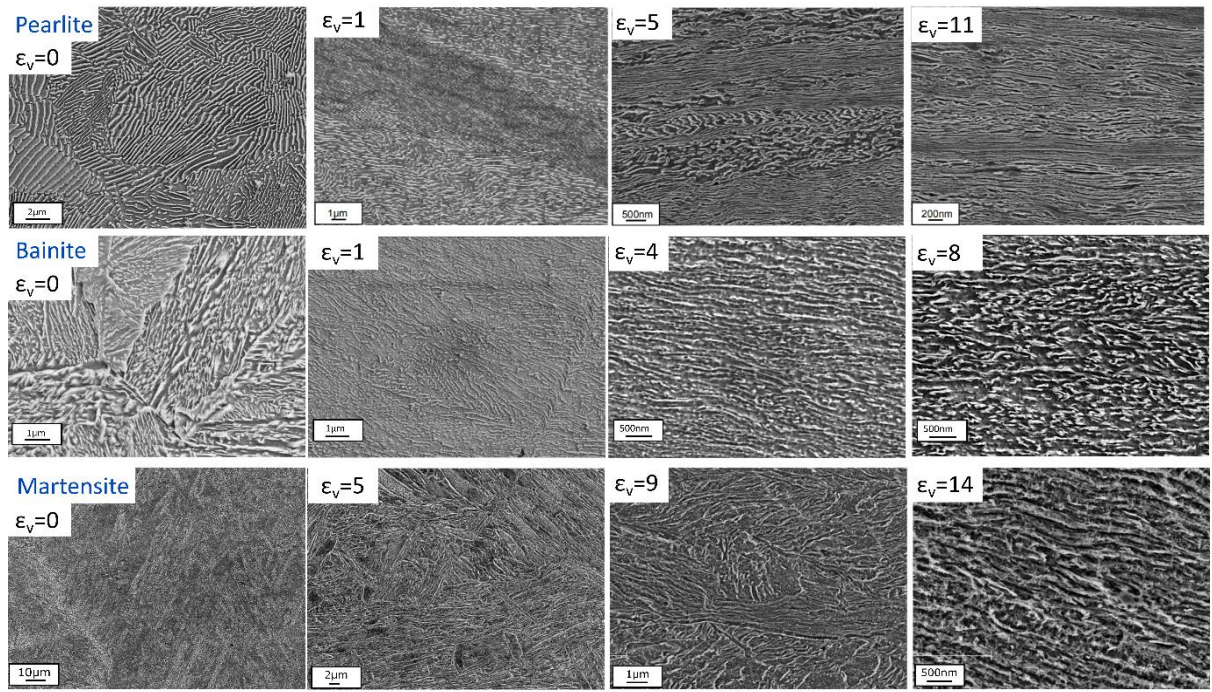

*Figure 3 Structural evolution during HPT with increasing equivalent strain, $\varepsilon_v$, for different steel grades during high pressure torsion: pearlite (R260, first row,[44]), bainite (Dobain, second row), martensite (Ck10, third row).*

Despite these similarities in the deformation procedure the final structure of HPT-processed pearlitic steels remains a nanocomposite at drawing strains of $\varepsilon_v$ = 16 (Figure 4) instead of a nanograined ferritic steel. Thereby, transmission electron microscopy (TEM) images reveal a two-fold nature of the pearlitic structure. In regions, where the initial lamellar alignment parallel to HPT shearing direction enables continuous thinning of the cementite, a perfectly lamellar compound of ferrite and cementite with an interlamellar spacing of 10-15 nm develops (Figure 4a). Contrast changes suggest that the ferrite phase is potentially subdivided into subgrains at moderate strains of $\varepsilon_v$ = 1.2 (Figure 4d) and are most likely still present at higher degree of deformation, but unfortunately not experimentally accessible due to the limited resolution of focused ion beam. However, this substructure is significantly larger in length with respect to the lamellar thickness which should make the latter the strength controlling spacing. The second region consists of a slightly coarser structure, where elongated ferrite grains dominate (Figure 4b). This structure might evolve either from already initially imperfect regions or from former colonies in which the cementite is unfavorably aligned with respect to the shearing direction and rather breaks or plastically ruptures. Still, the scanning electron microscopy (SEM) images in Figure 3 suggest that the lamellar structure predominates, which is reflected also in a pronounced anisotropy of mechanical properties [38,41,45]. Unlike to drawn wires, during HPT a planar alignment of the structural features is observed perpendicular to the main deformation direction (e.g. in TAN view, Figure 4c).

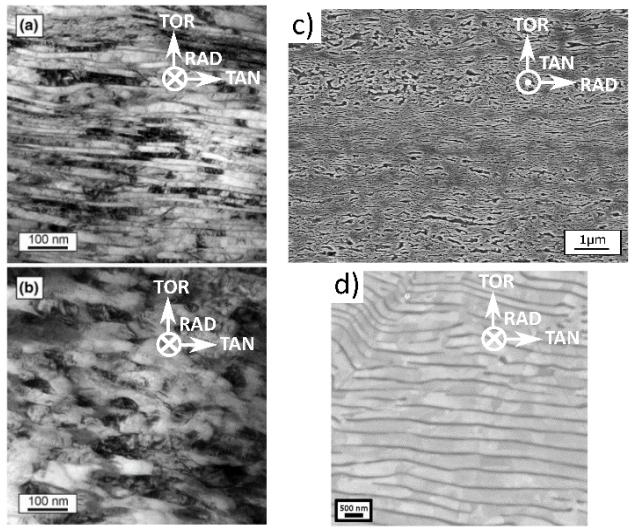

*Figure 4 TEM images of a pearlitic steel (R260) HPT deformed to $\varepsilon_v$ = 16 reveal structures of two fold architecture: a) perfectly lamellar arrangement of cementite and ferrite and b) elongated nanocrystalline ferrite grains when cementite is disrupted [41]. Reproduced with permission of Springer Nature ([Metallurgical and Materials Transactions A | Home (springer.com)](#)). c) Planar structural features develop also perpendicular to the shearing direction (SEM) and d) channeling contrast reveals subgrains within ferrite (FIB) for smaller strains of $\varepsilon_v$ = 1.2. TOR, RAD and TAN indicate the torsion, radial and tangential direction of the HPT process. Reprinted from [46] with permission from Elsevier.*

Electron energy loss spectroscopy reveals that the cementite becomes off-stoichiometric in these lamellar regions [43]. According to atom probe tomography (APT) analysis [47] the carbon concentration in the lamellae drops to values below 10 at% and gets redistributed to the ferrite matrix and its boundaries (Figure 5a). Thereby, the carbon concentration continuously decreases over the interface from the former cementite to the ferrite. This suggests, that processes such as dislocation drag or wear, initiating a continuous detachment of carbon atoms, govern the dissolution of lamellar cementite [47,48]. Yet, the non-stoichiometry of cementite increases with strain and it thereby transforms into a carbon-rich phase, that might have different mechanical properties as compared to the original cementite [45]. Earlier HPT approaches, using an unconstrained setup and thereby imposing nominal strains of $\varepsilon_v$ = 245 to a pearlitic steel [49], even report a complete dissolution of cementite according to x-ray diffraction (XRD) and thermomagnetic measurements. However, at the highest strains viable by quasi-constrained HPT ($\varepsilon_v$ = 16) only a moderate loss of carbon appears and a two-phase nanolamellar architecture prevails.

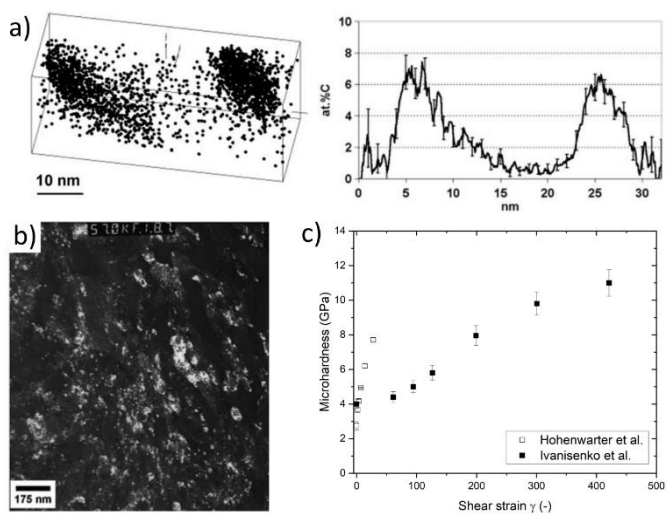

*Figure 5 a) APT on HPT-deformed pearlite (UIC 860V, $\varepsilon_v$ = 36, unconstrained setup) reveals loss of carbon in former cementite lamellae (e.g. carbon concentration < 25 at%) with gradual transition into the ferrite matrix [47]. Reproduced with permission from Springer Nature ([Journal of Materials Science | Home (springer.com)](#)). b) Ferrite grain sizes of 10 nm can be achieved at γ = 300 (UIC 860V, $\varepsilon_v$ = 173). Reprinted from [48] with permission from Elsevier. c) Microhardness of more than 10 GPa at γ = 430 (UIC 860V, $\varepsilon_v$ = 245, black squares) with an unconstrained setup, which is compared to hardness evolution with quasi-constrained setup (R260, open squares) from [41,49].*

Comparing the interlamellar spacing and the structural architecture at the maximum drawing strains of the different SPD techniques suggests different efficiencies in refining pearlitic steels. While for cold drawing comparatively smaller strains of $\varepsilon_d$ = 6.52 are sufficient to achieve ferrite subgrains with 8 nm spacing, during quasi-constrained HPT such small spacings cannot be achieved even at higher equivalent strains of $\varepsilon_v$ = 16. The ferrite spacing of roughly 15 nm and the moderate dissolution of the HPT structure correlates to drawing strains between $\varepsilon_d$ = 2 (minor dissolution, 30 nm ferrite spacing, Figure 1c) and $\varepsilon_d$ = 4 (distinct carbon enrichment in the matrix and at subgrain boundaries, 12 nm ferrite spacing Figure 1c) [15,17,50]. Even during unconstrained HPT nominal strains as high as $\varepsilon_v$ = 245 are necessary to achieve prevalence of a single phase structure, e.g. 10 nm ferrite grains with C stabilizing the interfaces (Figure 5b) [49].

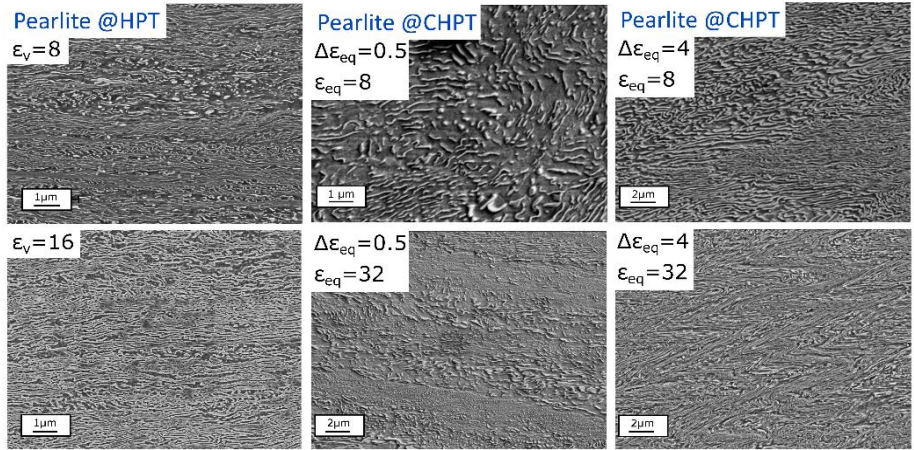

Figure 6 The nearly full lamellar alignment of pearlite (R260) at $\varepsilon_v$ = 8 and further refinement at $\varepsilon_v$ = 16 during HPT (first column) is not accessible by CHPT. While at small amplitudes $\Delta\varepsilon_{eq}$=0.5 ruptured carbides (second column, $\varepsilon_{eq}$ = 8) realign towards the HPT shearing direction ($\varepsilon_{eq}$ = 32), CHPT at higher amplitudes $\Delta\varepsilon_{eq}$=4 (third column) better resembles the monotonic HPT process, albeit a zig-zag structure remains.

The decelerated structural evolution as a function of strain is a consequence of differences in the strain path during wire drawing and HPT and partly of the different initial lamellar spacing. For co-deformation of the ferrite and the cementite during HPT, the lamellae spacing for the most efficient starting arrangement for refinement, e.g. cementite lamellae parallel to the torsion axis and the radial direction, at large shear strains, $\gamma$, is:

$$d = \frac{d_0}{\gamma} \tag{Eq. 3}$$

For wire drawing the most efficient lamellae arrangement for refinement is parallel to the drawing axis, and the spacing during can be calculated according to:

$$d = \frac{d_0}{e^{\varepsilon_d/2}} \tag{Eq. 4}$$

For this simplified assumption and for the considered cases a ferrite lamellae spacing of 4 nm and 8 nm could be estimated for wire drawing and HPT, respectively. The deviation by a factor of two is mainly caused from the initial random oriented lamellae within the colonies. The existence of a strain path dependency has been reported earlier [51] and is supported also by the structural evolution obtained during cyclic high pressure torsion (CHPT, Figure 6) [39]. In addition to differences in the interlamellar spacing the realignment of the structural features proceeds at slower rates during CHPT. While at $\varepsilon_v$ = 8 most of the lamellae are aligned along the shearing direction during quasi-constrained HPT, ruptured carbides ($\Delta\varepsilon_{eq}$ = 0.5, second column) or a zig-zag structure ($\Delta\varepsilon_{eq}$ = 4, third column), still dominate for CHPT, depending on the strain amplitude imposed per cycle. Especially for low cyclic amplitudes deformation of cementite is negligible, even at elevated strains of $\varepsilon_{eq}$ = 32. The lamellar refinement at large amplitudes better resembles the monotonic HPT structure, albeit a full lamellar alignment is still not accessible at $\varepsilon_{eq}$ = 32.

Thus, a refinement of the pearlitic structure to below 10 nm and the dissolution of the cementite phase, as observed for wire drawing is to date not accessible by HPT. One of the first HPT works in this field [49] observed a non-linear increase in hardness over shear strain and reported hardness values of more than 10 GPa at strains of $\varepsilon_v$ = 245 (Figure 5c). Yet, a one to one comparison with quasi-constrained HPT, causing linearly increasing hardness values (data from Figure 7b converted to GPa in Figure 5c for comparison), might not be straight forward with respect to the rate of carbon dissolution and structural evolution. Regarding deformation of such high-strength materials the possibility of slippage between anvil and sample always should be considered. Also for quasi-constrained HPT a structure more closely to the one achieved by wire drawing [15] or unconstrained HPT [49] would be accessible, if higher strains could be applied. However, the tool steel anvils restrict maximum deformability for quasi-constrained HPT to below 800 HV. This is clearly seen in the maximum achievable hardness values of the pearlitic steel grades, which do not exceed the anvil hardness (Figure 7) [44,52]. Once the deformed steels reach the hardness of the anvils, either slippage of the sample within the cavity of the HPT anvils or deformation of the anvil itself occurs. Thus, although the material might also get deformed beyond this hardness regime, the strict correlation between imposed HPT deformation angle and measured strain level gets lost. This should be always taken into account when HPT data are presented for a higher strains. Yet, from the current state an alternative anvil material could be a prospective solution, as for instance drawing dies either out of hard metal or polycrystalline diamond that provide high hardness and therefore enable high drawing strains.

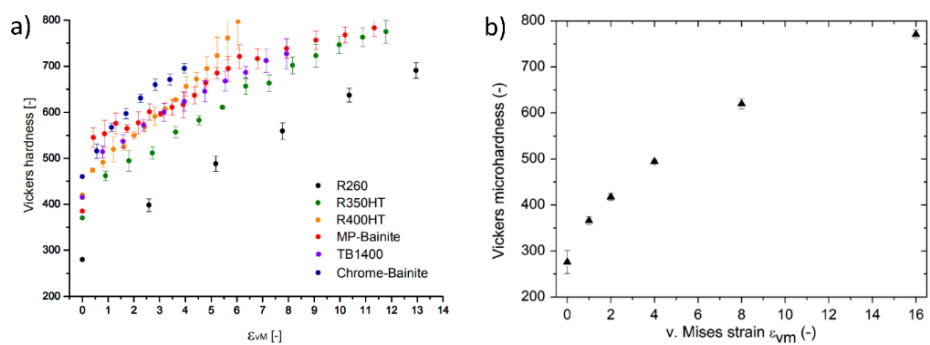

*Figure 7 Microhardness of HPT deformed a) pearlitic (R260, R350HT, R400HT) and bainitic (MP-Bainite, TB1400, Chrome-Bainite) steels grades up to HPT $\varepsilon_v$ = 14 [44,52] and b) the pearlitic steel R260 up to $\varepsilon_v$ = 16 [41]. Reproduced with permission of Springer Nature (Metallurgical and Materials Transactions A | Home (springer.com)).*

## 2.3 Structural evolution of quenched steels

Although HPT deformed high carbon steels deformed with conventional steel anvils cannot reach the outstanding strength of 6.8 GPa of severely drawn pearlitic steels with the quasi-constrained setup, the HPT technique is applicable to a wider range of steel grades. In contrast to cold drawing of wires, HPT does not rely on an optimized starting structure providing sufficient ductility to prevent cracking of the wires. Instead, steel grades with limited deformation capabilities, such as bainitic and martensitic steels, can also be processed. This is possible by the hydrostatic pressure, which suppresses large scale cracking in the samples up to a certain degree of deformation.

As an example, Figure 3 shows the deformation of a lower bainitic steel, initially consisting of ferrite plates and carbide precipitates[53]. The bainitic plates are a couple of microns wide and the carbides are evenly distributed within the ferrite plates but randomly in the overall structure (Figure 3, second row, $\varepsilon_v$ = 0). The ferrite plates deform uniformly, forming a lamellar shape similar as observed in the pearlitic steel. However, the carbides fragment ($\varepsilon_v$ = 1), realign along the shearing direction of the HPT disc ($\varepsilon_v$ = 4), but their thickness does not refine considerably ($\varepsilon_v$ = 8). Continuous thinning is inhibited by their larger size and their discontinuous nature compared to the cementite lamellae of pearlite, preventing co-deformation of the two phases. Nonetheless, the final structure is constituted by nanostructured ferrite grains lengthened along the HPT shearing direction and delineated by a carbon phase, which is very similar to drawn pearlitic wires despite their different starting architecture. Also the maximum achievable hardness does not exceed 770 HV (Figure 7a), thus further refinement and strengthening is restricted again by the tool steel anvils.

Deformation of a quenched lath martensite is also possible by tool steel anvils as long as the carbon concentration is low enough. For the Ck10 martensitic low carbon steel, consisting of ferrite laths being 200 nm in width, a similar structural evolution as shown for bainitic steels takes place (Figure 3, third row $\varepsilon_v$ = 0). This is a consecutive realignment of the initial lath structure towards the shearing direction of the HPT disc ($\varepsilon_v$ = 5), followed by a refinement of the already aligned ferrite laths ($\varepsilon_v$ = 9) and a transformation into a lamellar ferrite structure ($\varepsilon_v$ = 14). The fully aligned ferrite structure revealed by TEM at $\varepsilon_v$ = 15 (Figure 9) is identical to the lamellar regions of HPT deformed pearlitic steel grades (compare Figure 4a) [54], besides the absence of a delineating, continuous carbon-rich film. Nevertheless, the nanolamellar structure is stabilized by carbon at the grain boundaries.

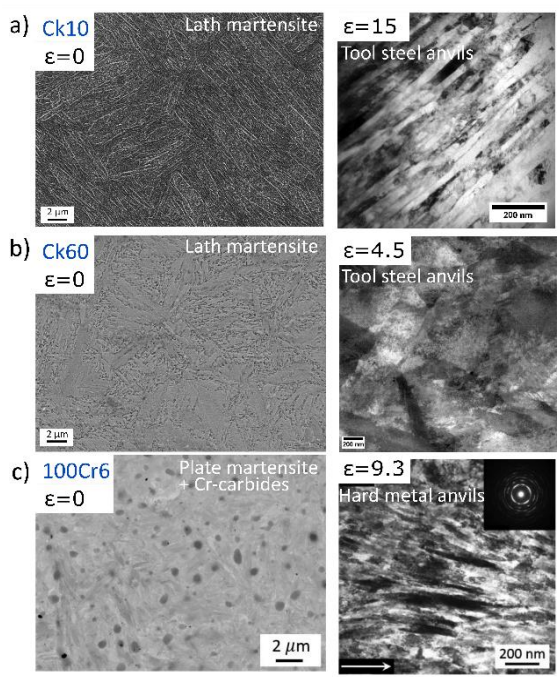

*Figure 8 SEM and TEM images of the structure before (first column) and after HPT deformation (second column) of different martensitic steels: a) low carbon steel Ck10, reprinted from [54] with permission from Elsevier; b) medium carbon steel Ck60 [55] and c) high carbon steel 100Cr6, reprinted from [56] with permission from Elsevier.*

Usually, HPT deformation of single phase materials yields such elongated grains only if boundary migration processes, sustaining a more globular grain shape, are suppressed under specific deformation temperatures [57,58]. As HPT was carried out at room temperature, the formation of such a lamellar ferrite architecture is therefore only possible, because the boundaries are stabilized by the segregated carbon atoms [54]. APT measurements revealed that almost the entire carbon atoms, dissolved in the martensitic matrix in the as-quenched condition, were transferred to the ferrite grain boundaries by strain induced segregation. However, the realignment of the martensitic structure requires a higher degree of HPT deformation. In Figure 3a nearly full alignment is observed at $\varepsilon_v$ = 4 and 5 for the pearlitic and bainitic grades, respectively, while strains as high as $\varepsilon_v$ = 14 are required for the martensitic steel. Probably, the need to transfer carbon atoms dissolved in the matrix from regions inside the ferrite laths to the boundaries by solute drag of dislocations [34] requires higher strain levels compared to the realignment of cementite and carbide in pearlite and bainite, respectively. Still, it is notable that small carbon concentration of only 0.1 wt% are sufficient to synthesize a mean ferrite lamellae spacing of only about 30 nm yielding 750 HV at strains of $\varepsilon_v$ = 15 (Figure 9a). This is about one order of magnitude smaller as the minimum structural size obtained in ARMCO iron (200nm subgrains [59] and 310nm grains [60]). Thus, also with such lean-alloyed steels the critical hardness limit inherent to deformation with tool steel anvils, as usually observed for eutectoid steels, is attained.

It should be mentioned that deformation of inherently more brittle materials like high-carbon martensitic steels is possible and yields exceptional strength values [54,61], however, it goes along with crack formation at elevated strains. Deformation to even higher strain levels is therefore impeded not only by the hardness of the tool steel anvils but also by deformation-induced cracking, often related to stress relief during unloading. Based on the results of the Ck10, deformation of martensitic steels with even higher carbon concentrations would promise further strengthening, in case that the additional carbon-content stabilizes an even finer microstructure [62]. Indeed, a hardness of 1025 HV is achieved at $\varepsilon_v$ ~ 32 after HPT with 0.6 wt% carbon (Figure 9b at $r$ = 3 and 720°), but the strengthening upon deformation, e.g. increasing deformation angle, is not pronounced. This is a consequence of the already initial high hardness of 825 HV for the Ck60 in the as-quenched state (Figure 9b). Unlike the Ck10, exhibiting only 425 HV (Figure 9a), the Ck60 exceeds the hardness of the tool steel anvil already prior to deformation. A realignment and consecutive refinement of the lath structure is therefore not pronounced (Figure 8b), as rather slippage between sample and anvil or deformation of the anvil itself takes place.

For the observed hardness increase of the Ck60 by 200 HV two points are assumed to be essential. First, smaller HPT discs (diameter of 6 mm for Ck60 instead of 8 mm for Ck10) enabled a higher hydrostatic pressure. Slippage and propensity for crack initiation, usually inhibiting further deformation of the structure, could thereby be retarded. Second, the Ck60 contains in the as-quenched state of about 5% of retained austenite. Due to the softer nature of this phase it could be further deformed (e.g. at least substructure formation). In addition, the austenite is transformed into martensite during deformation, thus, contributing to the hardness increase by a deformation-induced phase transformation. Yet, more investigations are required to analyze the contributions of these two effects.

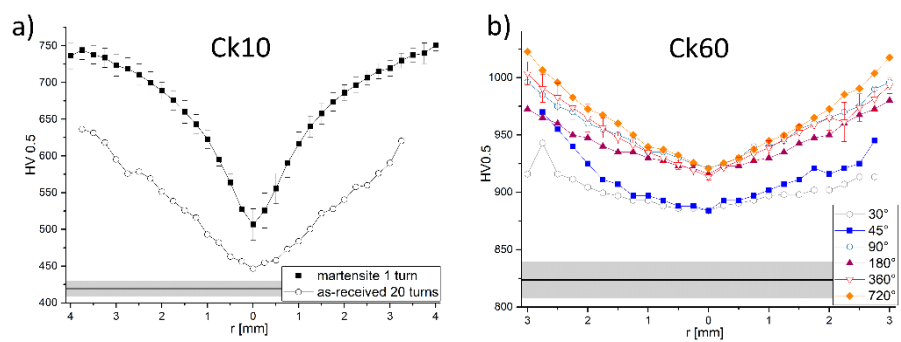

Figure 9 Microhardness along the HPT disc radius of the martensitic steels a) Ck10 and b) Ck60 [55]. The gray bars indicate the hardness of the as-quenched state.

Even if high hydrostatic pressure prevents slippage and crack formation, the applied strain is split into deformation of the sample and the anvils in an uncontrolled way if the strength of the sample is equal or even larger than the strength of the anvils. Well-defined HPT deformation of martensitic grades with elevated carbon-content is therefore only possible, when the hardness of the anvil is considerably higher than the starting material to ensure its deformation. A prospective solution is the use of hard metal anvils. This is shown for a 100Cr6 steel, deformed by HPT after oil quenching and subsequent tempering, leading to a plate martensitic matrix with homogenously distributed spheroidal cementite [56]. Compared to the Ck10 the starting structure is significantly coarser, as the plates have thicknesses between 160 nm and 1.2 µm. Despite a high initial hardness of roughly 825 HV and usage of a quasi-constrained setup the matrix refines with similar efficiency (Figure 8c) as the martensitic Ck10 (Figure 8a) and pearlitic R260 (Figure 4a) steel grades. An architecture with nanolayered ferrite grains of 20-40 nm spacing is achieved [56], yielding a hardness of 1050 HV after only 0.75 revolutions or $\varepsilon_v$ = 9.3 [63]. Thus, usage of hard metal anvils is a feasible strategy to overcome the current hardness limits of HPT deformed steels.

## 2.4 Strengthening mechanisms in nanolamellar steels
### 2.4.1 Hardness evolution with shear strain

Altogether, the structural evolution of steels during HPT yields in a lamellar architecture (Figure 10) with carbon playing a crucial role in stabilizing ferrite interfaces [52]. Yet, the efficiency to achieve structural refinement and so high strength depends on the steel grade that varies in starting architecture (lamellar cementite vs. random carbides), initial grain spacing or chemical composition. This also complicates a straightforward comparison of the different synthesis routes (e.g. wire drawing, unconstrained HPT, quasi-constrained HPT) as the discussed steel grades also differ in chemical composition (compare Table 1).

According to hardness measurements (see Figure 7a) starting architecture has the larger impact on the hardness evolution than chemical differences. The chemical differences in conjunction with the processing parameters are significant for the starting structure. The hardness of all bainitic grades not only surpasses the pearlitic steels at the same strain level, but also attains the anvil hardness (800 HV) at lower strain values. This might either be related to the easier alignment of the shorter carbides in lower bainite (no co-deformation of ferrite-cementite required) and thus, to a more efficient refinement of the ferrite grains, or the coarser cementite-carbides that prevail in the final structure additionally contribute to the hardness (Figure 3, second row).

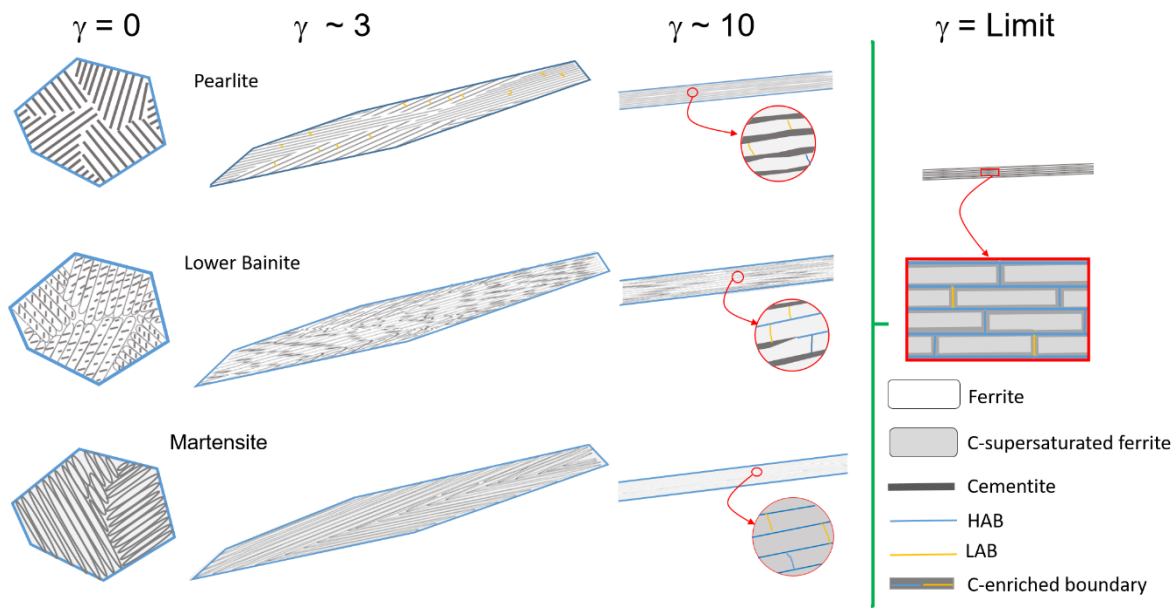

*Figure 10 Schematic illustration of the structural evolution of pearlitic, bainitic and martensitic starting materials during quasi-constrained HPT up to its current limits.*

With focus on pearlitic steels, a smaller initial interlamellar spacing is a further source for improved strain hardening. The hardness of the finer pearlitic grade (120 nm, see R350HT in Figure 7a) attains the anvil hardness already at $\varepsilon_v$ = 12 (Figure 7a), while the same hardness level requires considerably higher strains of $\varepsilon_v$ = 16 for a coarser starting structure (250 nm, see R260 in Figure 7b) [44]. Assuming that confined plasticity within the lamellae largely controls the hardness, efficient refinement of its spacing could explain this behavior. It relates well to earlier results, revealing that cementite deformability improves with decreasing thickness, thus, interlamellar spacing [12]. The fine cementite may accelerate the realignment of unfavorably aligned colonies and thereby promotes the overall refinement of the composite.

The impact of alloying on the hardening behavior of pearlitic structures can be seen in the comparison of the R350HT and R450HT grades in Figure 7a. Both have an initial lamellar spacing of about 110 nm, but R450HT contains higher carbon, chromium and manganese concentrations. The enhanced alloying gives rise for another hardness boost to 750 HV at $\varepsilon_v$ = 5 (Figure 7a). Also the strain level at which the maximum hardness is achieved drops to strains below $\varepsilon_v$ = 6, which is only one third of the strain necessary to achieve the same hardness with the standard grade (R260). The hardness increase stems thereby from a higher overall cementite fraction, that may reduce the initial lamellar spacing, due to elevated carbon and chromium concentrations, while manganese mainly contributes by solid solution hardening [44,64].

*Table 1 Differences in the chemical composition of SPD deformed steels presented in this review article (except for R400HT, MP-Bainite, TB1400, Chrome-Bainite due to confidentiality of its chemical composition).*

| Structure | Steel grade | Refs. | Composition | SPD Process |
|---|---|---|---|---|
| Pearlitic | Not specified | [15,27] | 0.98 C, 0.31 Si, 0.20 Mn, 0.20Cr | Wire Drawing |
| | R260 | [37–39,41,43–45,65] | 0.62-0.80 C, 0.15-0.58 Si, 0.70-1.20 Mn, ≤0.15 Cr | HPT, CHPT |
| | R350HT | [44] | 0.72-0.80 C, 0.15-0.58 Si, 0.70-1.20 Mn, ≤0.15 Cr | HPT |
| | R400HT | [44] | 0.90-1.05 C, 0.20-0.60 Si, 1.00-1.30 Mn, ≤0.30 Cr | |
| | UIC860V | [47,49] | 0.60-0.80 C, 0.10-0.50 Si, 0.80-1.30 Mn | |
| Bainitic | Dobain | [39] | 0.20-0.50 C, 0.80-1.20 Si, 1.20-1.80 Mn, ≤0.40 Cr | |
| Martensitic | Ck10 | [54,55] | 0.07-0.13 C, ≤0.40 Si, 0.30-0.60 Mn | |
| | Ck60 | [55] | 0.57-0.65 C, ≤0.40 Si, 0.60-0.90 Mn, ≤0.40 Cr, ≤0.40 Ni | |
| | 100Cr6 | [48,56] | 0.93 C, 0.33 Si, 0.41 Mn, 1.54 Cr, 0.11 Ni | |

Nevertheless, high hardness and strength values are also accessible for lean alloyed steels. For instance, despite the low carbon content of only 0.1 wt% in a Ck10 strain hardening is limited by the anvil hardness (Figure 9a) similar to its eutectoid pearlitic counterpart (R260 in Figure 7b). This means, that an eight times lower carbon concentration as compared to the pearlitic steel is sufficient to stabilize a ferrite grain structure of 30 nm (Figure 8), capable of reaching 750 HV microhardness at strains of $\varepsilon_v$ = 15 (e.g. *r* = 4 mm in Figure 9a). Carbon segregation at the interfaces, but no cementite precipitation is found by XRD and APT measurements after HPT [54]. Yet, an increase in the carbon concentration by a factor of six in the Ck60 further boosts the hardness to a maximum of 1050 HV (Figure 9b). However, if this is related to a further refinement of the structure [62], solid solution hardening [49] or a stronger distortion of the laths leading to increased internal stresses, has not been studied so far. As clearly visible from the only moderate hardness increase with increasing strain, strain hardening is not efficient anymore due to experimental restrictions. As mentioned before, when the hardness of the anvil material (tool steel, about 800 HV) is exceeded, slippage and anvil deformation minimizes further structural refinement in the sample.

## 2.4.2 Strengthening mechanisms in lamellar architectures

It is still notable that despite the different carbon-level and phase distribution, the martensitic Ck10 and the pearlitic R260 attain similar anisotropic strength levels [45,54,66]. This raises the question on the nature of hardening contributions to such lamellar structures. While for nanopearlitic steels in the past mainly the ferrite grain spacing was considered (compare Eq. 1 in [12]), earlier approaches sum up the individual hardening contributions to derive the flow stress *σ(ε)* with increasing strain [13]:

$$\sigma(\varepsilon) = \sigma_0 + \sigma(b) + \sigma(\rho) + \sigma(ss) \qquad (Eq.\ 5)$$

Boundary strengthening, $\sigma(b)$, increasing dislocation density $\sigma(\rho)$ and carbon in solid solution, $\sigma(ss)$, to the ferrite friction stress, $\sigma_0$, contribute to the total strengthening effect. However, according to anisotropic deformation experiments on such pearlitic nanocomposites [45,54] this approach might be too simplified. A straight forward comparison of different lamellae spacings by micro compression experiments emphasized the importance of dislocation confinement for strengthening [45]. This is further supported, when considering that deformation mechanisms in these nanostructures rely on emission and accommodation of dislocation within the interfaces [67–69], followed by movement of half loops through the lamellar channel in elongated grains. The latter is known as confined layer slip and is a well-studied feature in multilayers [17,70–75] causing an exceptional strain hardening capacity [76] and therefore ductility [38] in such architectures.

Confined layer slip also controls the flow stress in uniaxial experiments. Aligning the elongated structure under an angle inclined to the imposed load, e.g. lamellae parallel to maximum shear stress, $\tau$, yields a flow stress that is reflected by the Orowan stress, $\sigma_{or}$, with the shear modulus in ferrite, $G$, the burgers vector in ferrite, $b$, Poisson's ratio, $v$, and ferrite lamellae spacing, $d$, similar to [76,77]:

$$\sigma_{or} \sim 3\tau = \frac{Gb}{\pi(1-v)d} \ln \frac{d}{2\pi b} \qquad \text{(Eq. 6)}$$

Strength levels of 2.8 GPa are achieved in nanopearlitic steels ($d$ = 15-20 nm) deformed to $\varepsilon_v$ = 14 by quasi-constrained HPT [45]. For this specific loading situation, deformation of the cementite is not required, as the overall compression of the sample is realized by localized slip in ferrite. However, this is also the weakest loading direction of a lamellar structure. When the lamellae are aligned perpendicular or parallel to the imposed load, the elastic response of the non-stoichiometric cementite enables strain hardening within the ferrite up to way higher strength levels of roughly 3.5 GPa. This is a fingerprint of nanocomposites [78], when the lamellar spacing below 50 nm [70] favors motion of dislocation half loops along the channels. For pearlitic grades in particular, the crucial role of the interface for plasticity was highlighted in MD-simulations [79–81], revealing that the interface type controls not only the type of emitted dislocations, how they contribute to slip transfer from ferrite to cementite, but also their accommodation in the interface governs strength and ductility. Recent approaches aim to unite confined layer slip theory with the classical Hall-Petch concept by directly correlating the flow strength, $\tau$, to the dislocation density stored in the interface, $\rho_{int}$, [82]

$$\tau = \alpha G b \sqrt{\frac{\rho_{int}}{d}} \qquad \text{(Eq. 7)}$$

Yet, it is interesting to note that the prevalence of a second harder phase might not be required to observe anisotropic deformation characteristics in nanolamellar structures. Despite the completely different interface structure in the HPT-deformed Ck10 (i.e. carbon-enriched grain boundaries instead of continuous lamellar second phase), compression of an inclined structure yields also lower yield strength as compared to parallel and perpendicular loading [54]. This implies that a second phase and its concomitant elastic response might not be necessary to achieve anisotropic strength levels, when the grain boundary provides sufficient resistance for dislocation absorption or transmission. Nevertheless, the reduced strain hardening capacity as well as stronger tendency for strain localization in the nanolamellar Ck10 (30nm grain thickness, single phase, carbon-stabilized interfaces) and also in the nanograined pearlitic wire (8 nm subgrain size, single phase, carbon-stabilized interfaces) in contrast to the nanolamellar pearlite (15-20nm, ferrite-cementite interface) highlight the importance of a two-phase architecture to obtain ductility in ultrastrong structures [38].

These results imply that a deeper understanding of mechanical properties of nanolamellar structures requires extended efforts to reveal basic dislocation-interface processes [67,83]. How plasticity changes with the interface type (e.g. chemistry, misorientation, orientation relationships in case of composites) determines nanomaterials' properties but is also a clue how to tailor grain architectures in the absence of alloying elements [57]. This will not only allow to design strong nanomaterials with exceptional toughness [27,84] and ductility [38], but also to provide insights and solutions for steels that are prone to contact-fatigue issues like white etching layers during in service.

## 2.5 Conclusion

SPD of nearly fully pearlitic, bainitic and martensitic steels is summarized in this paper with a focus on HPT and wire drawing, because of their particular suitability to deform materials of high hardening capacity. The following conclusions can be drawn:

- » The strongest steels accessible to date are cold drawn pearlitic steels.
- » A similar structural evolution is obtained during HPT and wire drawing of pearlitic steels. However, HPT requires significantly larger equivalent strains to achieve the same structural sizes and degrees of carbon-dissolution as compared to wire drawing.
- » The strength of the anvil materials and available hydrostatic pressure is currently a major limit in HPT deformation of these steels.
- » SPD deformation of different steel types yields in a similar final architecture, which is elongated ferrite with carbon stabilizing its grain boundaries. Yet, for pearlitic steels elevated strain levels are necessary for a transition from a nanolamellar to a nanograined structure.
- » Similarities in anisotropic mechanical properties of SPD deformed pearlitic and martensitic steels highlight the importance of grain architecture for mechanical properties. However, the role of the interface in controlling deformation, strength and fracture is similarly important and requires increasing scientific attention in future studies.

Acknowledgement

Financial support by the Austrian Academy of Sciences via the Innovation Fund project IF 2019–37 and the Austrian Science Fund FWF within project number T 1347-N is gratefully acknowledged. This project has received funding from the European Research Council (ERC) under the European Union's Horizon 2020 research and innovation programme (Grant No. 757333).